# DB InfraGO's Automated Dispatching Assistant ADA-PMB

Stephan Zieger[1], Hannah Richta[1]
[1]DB InfraGO AG, Frankfurt, Germany
E-mail: stephan.zieger@deutschebahn.com, hannah.richta@deutschebahn.com

**Abstract**
As railway infrastructure manager, DB InfraGO AG is faced with the challenge of offering fluid and punctual operation despite rising demand and increased construction activity. The high capacity utilisation, especially in the core network sections, causes delays to be propagated quickly and widely across the entire network. Up to now, conflicts between train runs can be identified automatically, but dispatching measures have been based on past human experience.

An automated dispatching assistance system is currently being piloted to provide support for train dispatchers in their work. The aim is to offer them helpful dispatching recommendations, particularly in stressful situations with a high conflict density in the network section under consideration, in order to ensure the most efficient operation of the system.

The recommendations are currently displayed separately alongside the central control system. In future, they will be integrated into the central control system, which will significantly simplify communication between the train dispatcher and signal setter. Further development steps for the integration process are also presented and discussed.

## 1  Introduction

The political ambition to shift a large proportion of passenger and freight transport to rail (European Commission, 2024) means that, on the one hand, the strain on the core networks is growing and, on the other, passengers and companies expect adequate service provision. As the infrastructure often cannot grow in line with demand at ease, it is necessary to maintain stable operations through proper scheduling. When a large portion of the train paths that can be constructed is utilised due to the high demand, the buffer times in the timetable are often minimal and higher delays are often transferred (Zieger, Weik, & Nießen, 2018).

At the infrastructure manager, the signal setter and train dispatcher are responsible for keeping to the timetable, stabilising operations and returning to the timetable in the event of deviations. The signal setter is responsible for the direct operational safety-related business in a usually compact and clearly demarcated area. Dispatchers, on the other hand, often monitor corridors and longer sections of track or larger networks and coordinate the actions of the signal setters, as shown in Figure 1.

Up to now, dispatching decisions have often been made based on past experience or according to a scenario-based scheme. The large number of conflicts makes it all the more difficult to identify good conflict solutions in general, some of which also require dependent chains of action. Furthermore, communication between train dispatchers and signal setters is usually by telephone, resulting in additional work for each dispatching adjustment.

One approach to a solution is an automatic dispatching assistance system that helps train dispatchers with their work in real time by suggesting solutions for all foreseeable conflicts in the monitored area. Ideally, this assistance is integrated into the infrastructure manager's central control system to enable digital communication between train dispatchers and signal setters, with telephone communication only taking place in the event of enquiries regarding decisions based on better local knowledge. Such an assistance system is currently being piloted by the German infrastructure manager DB InfraGO AG and is presented in this paper.

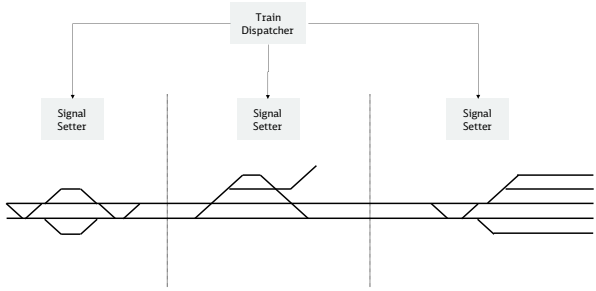

Figure 1: Area Responsibilities of Train Dispatchers and Signal Setters

## 2 The Automated Dispatching Assistance ADA-PMB

Train dispatching assistance systems have been a subject of research for decades. Section 2.1 provides a brief overview of the different approaches that have been pursued.

In Section 2.2 the model choice is discussed and in Section 2.3 ADA-PMB ('Automatische Dispositionsassistenz auf Basis Produktionsmodell Betrieb', Automatic Dispatching Assistance based on Production Model Operations) is presented. The previous publications (Blome, 2022; Rittner, Richta, & Große, 2022; Richta, 2023; Richta, 2024; Richta, Bermuth, Eversmann & Metzler, 2024; Wasserrab, Richta, Schwing & Helmrich, 2024) are not yet available in English, so a concise summary of the programme is provided. ADA-PMB uses the approach of solving an optimisation model and thus deriving near real-time operational stabilisation recommendations for the train dispatcher.

### 2.1 Real-Time Train Dispatching

Real-time train dispatching has been the subject of railway research for a long period of time. Schaefer and Pferdmenges (1994) provide a comprehensive overview of the challenges that need to be solved, whereby elements in parentheses are not considered by ADA-PMB, partly due to unavailable data:
- Track occupancy conflicts – the same track section is planned to be used by two trains
- Schedule conflicts – a train deviates from its planned schedule
- Closed-track conflicts – a track section is unavailable in contrast to its planned train path
- (Roster conflict) – Identical rolling stock or staff is used in separate journeys

The modelling options and a range of possible solutions are presented in D'Ariano (2008). A solution is in general a mix of train scheduling, rerouting and speed profile adjustments. Heuristics of different quality and accuracy (D'Ariano, 2008) or optimisation approaches (D'Ariano & Pranzo, 2008) are usually used to obtain solutions. Machine learning approaches based on historical data are also considered for decision-making (Yue, Jin, Dai, Feng, & Cui, 2023). Reinforcement learning approaches based on a traffic simulation are e.g. tested by DB Regio, the DB Groups regional train provider (Deutsche Bahn AG, 2023).

### 2.2 Method Choice for ADA-PMB

All of the methods mentioned have their advantages and disadvantages. Machine Learning based approaches have the disadvantage that they may come to good solutions, but they are mainly a black box, i.e., it is hard to explain why the algorithm found a certain solution. It is also hard to prove that this solution is the optimal solution. Therefore, particularly for DB InfraGO as a rail infrastructure manager who is obliged to not discriminate between the railway undertakings, those approaches seem not feasible.

Most heuristics solve two-train conflicts one after the other, which are sorted in a suitable manner - for example chronologically or according to a calculated conflict severity. The solutions can lead to further subsequent conflicts, which must also be resolved. The approach is often based on a strict set of rules that are intended to derive on average good but local decisions. This results in a further disadvantage of the approach, namely that it is difficult to measure the degree of success of the approach.

The size of solvable mathematical optimisation models has increased rapidly over the last few decades. Nevertheless, the models usually grow superlinearly, so even with powerful hardware, the possibilities for solving the problem remain limited (cf. Section 3.2). However, the optimisation approach also offers advantages. First, conflicts are not only solved locally, but globally (in the sense of the defined spatial and temporal area under consideration), and second, the gap between best currently found and best known bound provides an objective measure for determining the distance to a verifiably optimal solution. Both points are particularly advantageous in terms of proving compliance with regulations.

To improve the speed of the solutions, modern commercial solvers accept hints. These hints point the solver towards a specific part of the solution space. While the solver is not bound to these hints, good hints are able to speed up the process of solving an optimisation problem. This results in the opportunity to mix the mathematical optimization with all its advantages with other approaches like heuristics and machine learning.

### 2.3 Workflow of ADA-PMB

The various components of ADA-PMB are highly complex and it would go beyond the scope of this document to describe them all. Therefore, the individual steps are only briefly and concisely described. A flow chart of the programme is shown in Figure 2.

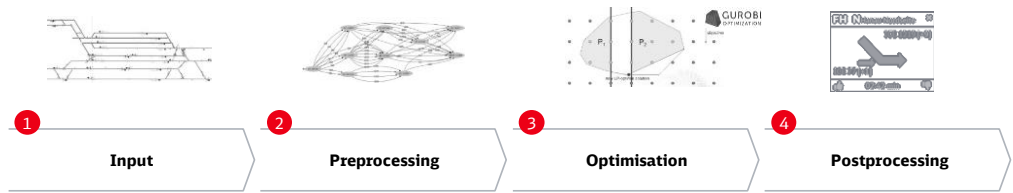

Figure 2: Workflow of ADA-PMB

**Input**
The input data is obtained from the operation centres. Initially, microscopic infrastructure data is obtained as the basis for all calculations, which contains both the topology and the positions of all elements which are relevant for the operation of train services, such as signals or stopping locations. Furthermore, availability restrictions are imported, in which the section in which a restriction exists is defined. In addition to the infrastructure data, the timetable and train data is also included. Thus, both the scheduled data and the actual data are available, including the location messages of the trains and the corresponding routes that have already been set by the interlocking.

**Preprocessing**
In the preprocessing step, a snapshot of the current operating situation is created. This is filtered to the spatial and temporal domain under consideration. The possible routes and speeds are determined for each train, as well as exclusions between routes. The complete pre-process is very technical, so only a minimum example is provided for the derivation of the possible velocity profiles (VProfiles) in Figure 3. At each main signal, for a train it must be possible to stop in the model. In addition, it must be possible to start up to maximum speed again from this point. Just these (non-exhaustive) possible combinations result in 10 VProfiles in the example for the section under consideration between the two main signals.

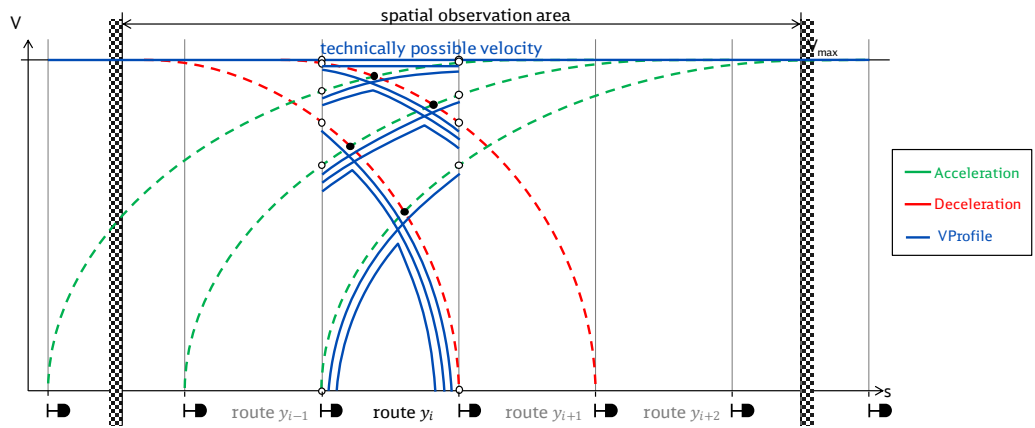

Figure 3: Creation of VProfiles for a single route considering different acceleration and deceleration possibilities

Additionally, to speed up the optimization process, hints are also created in the preprocessing step. This is necessary, since there is a tight time limit for each optimisation run. The hints are based on a mix of different methods. Lower bounds are calculated by applying heuristics. Then other heuristics check which parts of the previous optimization run's solution can be used as hints for the next one. The rationale behind this heuristic is that the traffic situation does not change completely between two optimisation runs and therefore, large parts of the previous solution may be still valid. So, the optimiser can save time by using these parts as hints. And finally, for trains for which the last solution cannot be used, a machine learning algorithm predicts the speed profile as hints. The last step is based on supervised machine learning, where previous traffic situations and the recorded optimisation results for these traffic situations are used to train an algorithm. The prediction of the speed profiles was selected because experiments had shown that these variables have a significant impact on the time needed for optimisation runs.

**Mathematical Model**

A mathematical optimisation model is created from the pre-processed information. The objective function is to minimise the overall delay at the customer stops in the area under consideration, taking into account the train priorities. The constraints include requirements for customer stops, running times, flow conservation, train sequence and other technical constraints for safe operation. The most relevant variables are the decision of the chosen path, the VProfile and the order of trains in a section. The model is in parts thus most similar to a flow problem on a event graph.

The blackbox solver by Gurobi (2024) is used to solve the model. The time to determine the solution is limited to 60 seconds in order to be as close to real time as possible. The solution is free of subsequent conflicts in the observation area and in line with the current regulations. Thus, the result is only postprocessed if it reached a certain gap, in the case of ADA-PMB the provable optimal solution must not be further away than 10%.

**Postprocessing**

The optimisation result is subsequently compared with the current status of the infrastructure and train allocations, as changes may have occurred between the snapshot and the current time. These are captured appropriately and the programme attempts to 'trace' the optimisation result, i.e. to adhere to the recommended speeds and times for arrivals and departures at operating points, taking into account the up-to-date information. Recommendations for the train dispatchers are then derived from the postprocess result.

The recommendations include changes in train order in all variations as well as track and line changes. Possible recommendations are shown as examples in Figure 4. The trains involved, the location and the time remaining for realisation are displayed. Train dispatchers can also use the thumbs up/down buttons to provide anonymous feedback on the recommendation. The recommendations are currently displayed in an additional window outside the central control system.

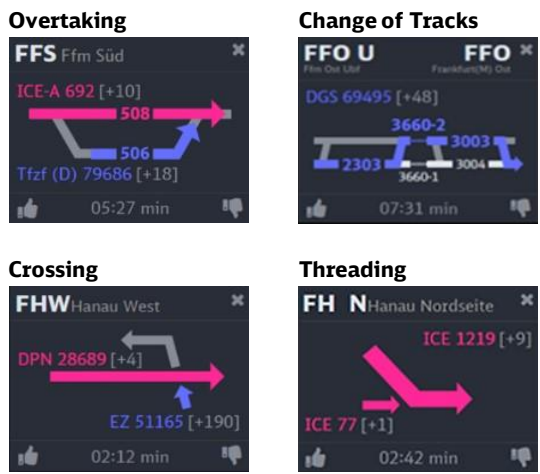

Figure 4: Different Types of Recommendations for Train Dispatchers

## 3 Path Towards the Integration of ADA-PMB in the Central Control System

Implementing an assistance system that is used on a large scale can hardly be done in a single step. Section 3.1 therefore presents the deployment strategy for ADA-PMB in order to make the programme technically ready for practical application on all relevant lines with their particularities. The individual deployment areas are foreseeably not stand-alone, which means that the areas must be properly defined and an exchange across the area boundaries is necessary. The ideas and details are outlined in Section 3.2. Finally, the current and future communication process between train dispatcher and signal setter is described in Section 3.3.

### 3.1 Deployment of ADA-PMB in Different Regions

The immediate network-wide implementation of a method such as ADA-PMB poses major uncertainties with regard to the performance of the programme and the acceptance of the assistance system. For this reason, ADA-PMB is initially being rolled out in stages in various network areas. On the one hand, rolling out the assistance system on individual lines identifies the need for further functional and technical development and, on the other hand, provides valuable feedback from end users. This approach also has advantages in the development process.

An overview of the active instances can be found in Table 1. The exact locations cannot be disclosed at this time and are thus summarised. The variation between the different regions is quite high with observation areas which consist only of pure urban traffic, but also mixed traffic railway lines are considered and supported. ADA-PMB featuring mixed traffic applications makes it special in the sense that a pure machine learning approach usually struggles with this kind of heterogeneity.

Table 1: Overview of the Active Instances of ADA-PMB

| Number | Train Journeys per Day | Urban Traffic | Local Traffic | Long-Distance Traffic | Freight Traffic |
|---|---|---|---|---|---|
| 1 | 1800 | x | | | |
| 2 | 500 | x | x | x | x |
| 3 | 2200 | x | x | x | x |
| 4 | 950 | x | x | x | x |
| 5 | 900 | x | | | |
| 6 | 500 | x | x | x | x |

### 3.2 Proper Sectioning of Observation Areas and Connection of the Same

As described above, it is not possible to calculate solutions for arbitrarily large observation areas. Therefore, there is always a trade-off between solution quality ('how often does the programme reach a defined gap in a predefined time?') and the size of the spatial and temporal observation area. The desired solution quality is usually defined, such as 'at least 90% of the optimisation runs should have a gap of no more than 10% in the peak traffic period'. A conflict solution for the next 20 minutes is often sufficient as a temporal constraint, as many factors can commonly change the train runs over a longer period. In some cases, however, a longer period is necessary if the topology requires it, i.e. single-track lines are considered, and the crossing stations are far apart.

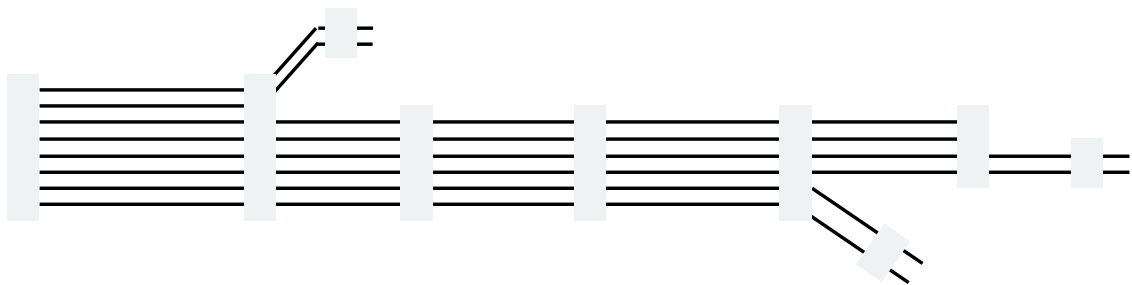

Figure 5: Sectioning of the Infrastructure in a Large Scale

Two limiting conditions must be considered for large-scale deployment. Firstly, each section on which the dispatching assistance is to be provided must be suitably sectioned and secondly, these areas must be linked in a suitable manner. Figure 5 shows a generic example of how the infrastructure can be divided into appropriate sections. The lines represent the tracks between the stations (boxes). Switches and signalling are omitted. Here, the search focuses on suitable boundaries where no more changes are made. In this example, these are the transitions to double-track lines.

At a more detailed level, it is not only important that the individual areas themselves operate effectively, but also that the respective results are exchanged with each other in a suitable manner. Clearly, the exchange of information is necessary for trains that cross the boundaries of the area under consideration, but this information can also have indirect effects for trains that only run within an observation area.

Figure 6 presents an example of the time-distance diagrams for two observation areas in the transition area. The solid line indicates the current prediction of the train run and the dashed line the optimisation result associated with the respective train.

Figure 6(a) depicts the initial situation with the three trains (red, blue and green), in which there are conflicts. On the one hand, the red and blue train have a conflict in Observation Area A and on the other hand, the green train has a conflict with the other two trains in Observation Area B. The train movements take place on the right hand side in the direction of travel.

A possible optimisation result is shown in Figure 6(b), where both observation areas do not exchange any information with each other. In observation area A, the conflict between the blue and red trains is resolved by overtaking. It is assumed that the green train enters into the area as planned. In Observation Area B, it is not known that a change in the sequence between the red and blue train can take place, but is taken into account as the prognosis says so. Accordingly, there is a crossing with both trains with a long waiting time for the green train and a significantly later breakout from the area.

Figure 6(c) displays the situation in which both observation areas exchange information on the respective optimisation results and these are incorporated into the next optimisation run. Only values from trains entering the area under consideration are used as input from other observation areas. There is no backward loop. The information about the sequence change between the red and blue train is now forwarded to observation area B allowing the crossing with the green train to be managed more effectively. The respective observation areas now no longer use the train run prediction, but rather the earliest possible incursions into the observation area, taking into account the dispatching measures, which can also have an impact on the calculation of further conflicts.

It is evident that the approach does not necessarily lead to global optima, but it does significantly improve the provision and quality of information. One obvious challenge here is the asynchronous runtimes of the two observation areas, which means that it may take several optimisation runs before a stable state is reached.

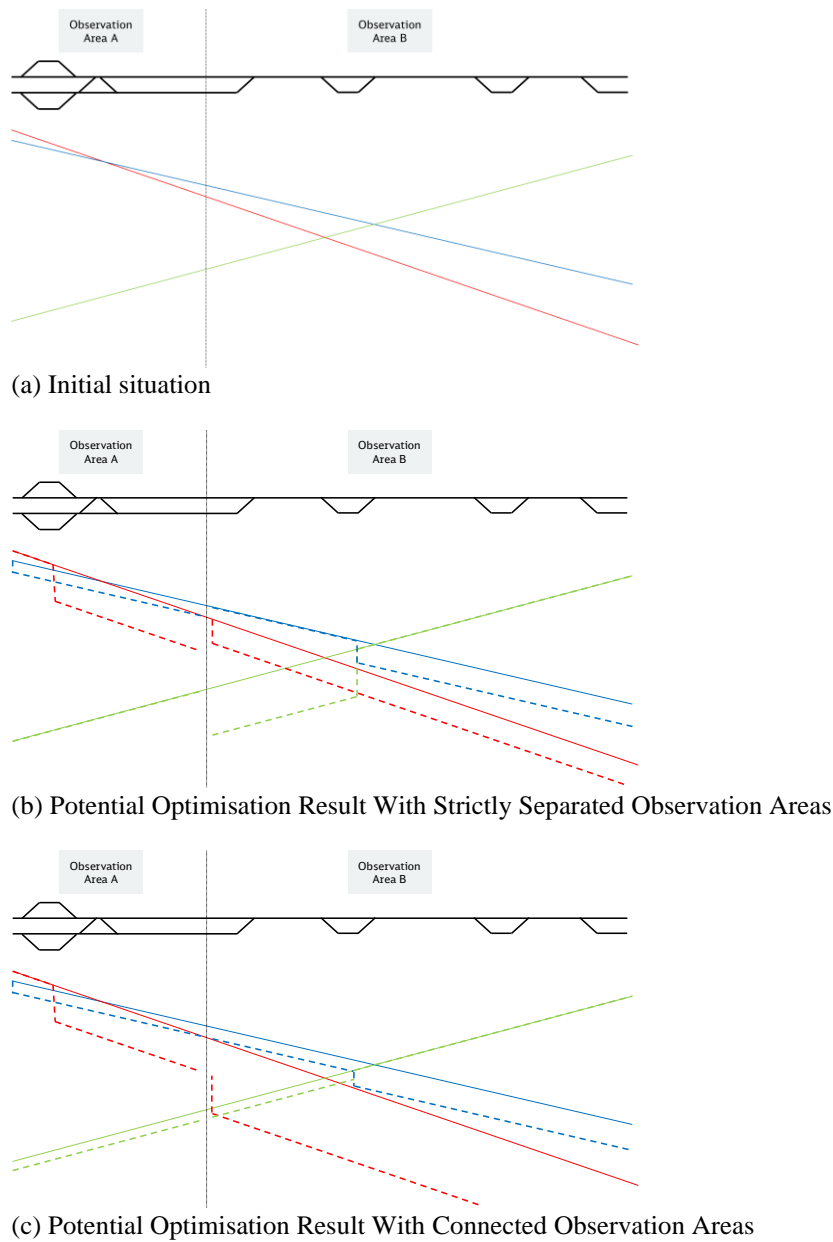

(a) Initial situation

(b) Potential Optimisation Result With Strictly Separated Observation Areas

(c) Potential Optimisation Result With Connected Observation Areas

Figure 6: Time-Distance-Diagrams of the Transition Area at the Border Between Two Observation Areas

### 3.3 Communication Between Train Dispatcher and Signal Setter

Currently, the train dispatcher is provided with the recommendations from ADA-PMB in an additional window, which can be positioned next to the central control system. For each recommendation, a train dispatcher first reads it from the window and then searches for the situation at the operating location with the corresponding train numbers in the central control system. If this recommendation is regarded as reasonable, a telephone conversation takes place with the locally responsible signal setter, who receives the instructions for realisation. This procedure is unsatisfactory from several perspectives and results in recommendations not being realised, especially in major stress situations with high conflict density.

 The aim is therefore to integrate ADA-PMB into the central control system in order not only to display the recommendations to the train dispatcher directly in the right place together with the effects of the change, but also to simplify communication. Figure 7 shows the process using the example of an overtaking of two trains.

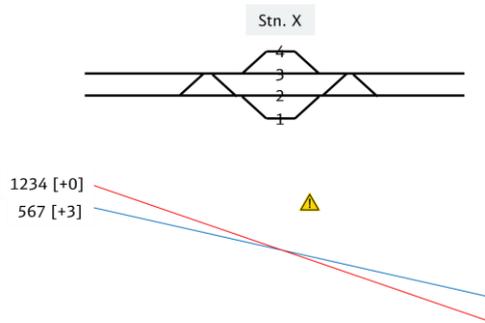

(a) Train Dispatchers View With Integration of ADA-PMB in the Central Control System

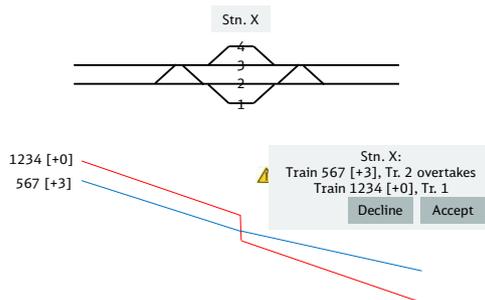

(b) Train Dispatchers View With Integration of ADA-PMB in the Central Control System After Hovering Over the Warning Sign

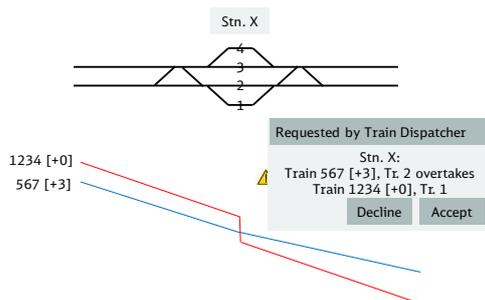

(c) Signal Setters View With Integration of ADA-PMB in the Central Control System After the Acceptance of the Train Dispatcher

Figure 7: Display of Recommendations With Integration of ADA-PMB in the Central Control System

Figure 7(a) abstractly represents a fragment of the central control system showing a conflict in station X between train 1234 with 0 minutes delay and the faster train 567 with 3 minutes delay. The warning symbol indicates to the train dispatcher that there is a conflict. In this situation, ADA-PMB recommends an overtake, which is displayed for the train dispatcher after clicking on the warning symbol. The dispatcher can now accept or reject the recommendation, as shown in Figure 7(b). If the recommendation is accepted, it is forwarded to the responsible signal setter as presented in Figure 7(c). The latter can then again accept or reject the recommendation. In other words, an additional conversation is only necessary if the train dispatcher and signal setter disagree.

## 4 Outlook

Liebchen and Schülldorf (2019) point out in their study that there are significantly more publications that aim to improve performance in the railway industry using mathematical optimisation approaches than actually successful implementations. With ADA-PMB, an assistance system based on mathematical optimisation has been developed, which has so far received a lot of positive, but also valuable critical feedback and is already having an impact in productive use. By extending it to other pilot areas, further feedback will be gathered and the underlying systems will be further optimised against edge cases.

    The railmap towards the implementation of ADA-PMB in the central control system is already very

well defined, but is contingent on the new central control system PRISMA (DB InfraGO AG, 2024) going live. Once PRISMA has been rolled out, further functions will be added. This also includes ADA-PMB, which will then be made available again in a staged process. Until then, the technical foundation needs to be strengthened and further steps, as described in Chapter 3, need to be taken.

**References**

Blome, A. (2022). Intelligentes operatives Kapazitätsmanagement bei der DB Netz AG. *Deine Bahn 10/2022*.

D'Ariano, A., & Pranzo, M. (2008). An Advanced Real-Time Train Dispatching System for Minimizing the Propagation of Delays in a Dispatching Area Under Severe Disturbances. *Networks and Spatial Economics*, 63-84.

D'Ariano, A. (2008). *Improving real-time train dispatching: models, algorithms and applications.* TRAIL Research School.

DB InfraGO AG. (2024). *PRISMA geht Live! Bestandssysteme werden im März 2025 abgelöst!* https://www.dbinfrago.com/web/aktuelles/kund-inneninformationen/kund-inneninformationen/2024-KW36-Prisma-geht-live-13067916# (checked 01.10.2024)

Deutsche Bahn AG. (2023). *Deutsche Bahn weitet Einsatz von KI für pünktlichere Züge aus*. https://www.deutschebahn.com/de/presse/pressestart_zentrales_uebersicht/Deutsche-Bahn-weitet-Einsatz-von-KI-fuer-puenktlichere-Zuege-aus-10771280 (checked 01.10.2024)

European Commission. (2024). *Mobility Strategy*. https://transport.ec.europa.eu/transport-themes/mobility-strategy_en (checked 01.10.2024)

Gurobi Optimization, L. L. C. (2024). Gurobi optimizer reference manual.

Liebchen, C., & Schülldorf, H. (2019). A collection of aspects why optimization projects for railway companies could risk not to succeed – A multi-perspective approach. *Journal of Rail Transport Planning & Management*.

Richta, H. (10 2024). Automatische Dispositionsassistenz ADA-PMB. *EI – DER EISENBAHNINGENIEUR*, S. 50-54.

Richta, H. N. (2023). Wie Künstliche Intelligenz helfen kann, die Pünktlichkeit der Bahn zu verbessern. In A. Gillhuber, G. Kauermann, & W. Hauner, *Künstliche Intelligenz und Data Science in Theorie und Praxis* (S. 293--30). Springer Spektrum.

Richta, H., Bermuth, K., Eversmann, D., & Metzler, T. (1 2024). Weiterer Ausbau der Automatischen Dispositionsassistenz. *Deine Bahn*, S. 50-55.

Rittner, M., Richta, H. N., & Große, S. (2022). Automatische Dispositionsunterstützung mit ADA-PMB. *Deine Bahn 10/2022*, S. 16-19.

Schaefer, H., & Pferdmenges, S. (1994). An Expert System For Real-time Train Dispatching. *WIT Transactions on The Built Environment.* WIT Press.

Wasserrab, F., Richta, H., Schwing, F., & Helmrich, V. (12 2024). Beschleunigung mathematischer Optimierung durch maschinelles Lernen. *Deine Bahn*, S. 30-34.

Yue, P., Jin, Y., Dai, X., Feng, Z., & Cui, D. (2023). Reinforcement learning for online dispatching policy in real-time train timetable rescheduling. *IEEE Transactions on Intelligent Transportation Systems*, 478 - 490.

Zieger, S., Weik, N., & Nießen, N. (2018). The influence of buffer time distributions in delay propagation modelling of railway networks. *Journal of Rail Transport Planning & Management*, 220-232.